\documentclass[useAMS,usenatbib]{mn2e}
\usepackage{graphicx, txfonts}
\usepackage{url}
\usepackage{multirow}
\usepackage{booktabs}

\usepackage[usenames,dvipsnames]{color}

\newcommand{\CI}{C\,\textsc{i}}
\newcommand{\CII}{C\,\textsc{ii}}

\newcommand{\CIII}{C\,\textsc{iii}}
\newcommand{\CIV}{C\,\textsc{iv}}

\newcommand{\DI}{\textrm{D}\,\textsc{i}}
\newcommand{\DII}{\textrm{D}\,\textsc{ii}}

\newcommand{\HI}{\textrm{H}\,\textsc{i}}
\newcommand{\HeI}{\textrm{He}\,\textsc{i}}
\newcommand{\HeII}{\textrm{He}\,\textsc{ii}}

\newcommand{\HII}{\textrm{H}\,\textsc{ii}}

\newcommand{\Lya}{Ly$\alpha$}

\newcommand{\NI}{N\,\textsc{i}}
\newcommand{\NII}{N\,\textsc{ii}}

\newcommand{\NIV}{N\,\textsc{iv}}

\newcommand{\OI}{O\,\textsc{i}}
\newcommand{\OIV}{O\,\textsc{iv}}

\newcommand{\MgI}{Mg\,\textsc{i}}
\newcommand{\MgIV}{Mg\,\textsc{iv}}
\newcommand{\SiI}{Si\,\textsc{i}}
\newcommand{\SiII}{Si\,\textsc{ii}}
\newcommand{\SiIII}{Si\,\textsc{iii}}
\newcommand{\SiIV}{Si\,\textsc{iv}}

\title[Deuterium Ionization Correction]
{The Primordial Abundance of Deuterium: Ionization correction}

\author[Cooke \& Pettini]{Ryan Cooke$^{1}$\thanks{email: rcooke@ucolick.org}\thanks{Hubble Fellow}, 
Max Pettini$^{2,3}$
\\
$^1$UCO/Lick Observatory, University of California, Santa Cruz, CA 95064, USA\\
$^2$Institute of Astronomy, Madingley Road, Cambridge, CB3 0HA\\
$^3$Kavli Institute for Cosmology, Madingley Road, Cambridge CB3 0HA\\
\\
}

\begin{document}

\date{Accepted . Received ; in original form }
\pagerange{\pageref{firstpage}--\pageref{lastpage}} 
\pubyear{2015}

\maketitle

\label{firstpage}

\begin{abstract}
We determine the relative ionization of deuterium and hydrogen in low metallicity
damped Lyman-$\alpha$ (DLA) and sub-DLA systems using a detailed suite of
photoionization simulations. We model metal-poor DLAs as clouds
of gas in pressure equilibrium with a host dark matter halo, exposed to the
\citet{HarMad12} background radiation of galaxies and quasars at redshift $z\simeq3$.
Our results indicate that the deuterium ionization correction correlates with the 
\HI\ column density and the ratio of successive ion stages of the most commonly
observed metals. The $N$(\NII)/$N$(\NI) column density ratio provides the most
reliable correction factor, being essentially independent of the gas geometry,
\HI\ column density, and the radiation field. We provide
a series of convenient fitting formulae to calculate the deuterium ionization correction
based on observable quantities. The ionization correction typically does not exceed
$0.1$ per cent for metal-poor DLAs, which is comfortably below the current measurement
precision ($2$ per cent). However, the deuterium ionization correction may need to be applied
when a larger sample of D/H measurements becomes available.
\end{abstract}

\begin{keywords}
methods: numerical -- galaxies: abundances -- galaxies: ISM --
quasars: absorption lines -- cosmological parameters -- primordial nucleosynthesis
\end{keywords}

\section{Introduction}

The relative abundances of the nuclides that were made just minutes
after the Big Bang currently provide our most reliable probe of
the physics and the content of the very early Universe. The most
well-studied of these primordial element abundances include the deuterium
abundance (D/H), the primordial $^{4}$He mass fraction (Y$_{\rm P}$),
and the $^{7}$Li abundance ($^{7}$Li/H).
If the standard model of cosmology and particle physics provides an
accurate description of the Universe throughout
Big Bang Nucleosynthesis (BBN), the primordial element abundances
depend only on the number density ratio of baryons to photons
($\eta_{10}=n_{\rm B}/n_{\gamma}$ in dimensionless units
of $10^{-10}$). If the ratio of baryons-to-photons is unchanged
from BBN to recombination, then BBN is a parameter-free theory
provided that $\eta_{10}$ can be measured with sufficient precision
from the Cosmic Microwave Background (CMB). Using the results from
the recent CMB analysis by the \citet{Efs15}, the primordial element
abundances for the standard model are predicted to have the values:
log\,(D/H)$_{\rm P}=-4.589\pm0.022$,
Y$_{\rm P}=0.24709\pm0.00025$, and
A($^{7}$Li)$_{\rm P}=12+\log(^{7}{\rm Li/H})_{\rm P}=2.666\pm0.064$
at 68 per cent confidence \citep{Cyb15}.

These predictions confirm the well-known
`lithium problem' (for a review, see \citealt{Fie11}); the standard model
$^{7}$Li abundance now constitutes a $4.3\sigma$ deviation
from the best observational determination,
A($^{7}$Li)$=2.199\pm0.086$, derived from the atmospheres
of metal-poor stars in the halo of the Milky Way \citep{Asp06,Aok09,Mel10,Sbo10,Spi15}.
At present, it is still unclear if new physics beyond the current standard model is
required during BBN to explain this discrepancy, or if some $^{7}$Li is destroyed
during the lives of metal-poor stars \citep[e.g.][]{Kor06}.

In addition to the lithium problem, there appears to be a `helium problem'.
Specifically, the primordial $^{4}$He mass fraction estimated by
\citet[][Y$_{\rm P}=0.2551\pm0.0022$]{IzoThuGus14}, deviates
from the standard model expectation at a level of $3.6\sigma$.
However, using a similar dataset and a different analysis strategy,
\citet{AveOliSki15} report Y$_{\rm P}=0.2449\pm0.0040$, which
is in agreeement with the standard model value. The discrepancy
between these two recent studies could be attributable
to the different sample selection cuts applied by these authors and/or
systematics that are presently unaccounted for (e.g. an
incomplete modelling of the emission line spectrum of the studied
\HII\ regions).

The primordial deuterium abundance reported recently by \citet{PetCoo12a}
and \citet{Coo14} is in good agreement with the standard model expectation.
The best environments to measure the primordial D/H abundance are the
most metal-poor damped Lyman-$\alpha$ systems (DLAs)
that are seen in absorption along the line-of-sight to a more distant
background quasar. Highly precise measurements of the D/H abundance
in these systems are made possible by:
(1) the simple and quiescent kinematic structure of the absorbing gas;
(2) the Lorentzian damped \Lya\ absorption wings, which depend sensitively
on the total column density of neutral hydrogen; and
(3) the host of weak, high-order Lyman series absorption lines
of neutral deuterium, whose equivalent widths are directly
proportional to the total column density of neutral deuterium.
The deuterium abundance then follows by simultaneously
fitting the relative strengths of the high order \DI\ absorption
lines and the \HI\ Lyman series lines (including the crucial
\Lya\ absorption feature). For these reasons, the measurement
of D/H is arguably the most reliable primordial element abundance,
since its determination is almost entirely independent of the modelling
technique employed.

Although the modelling technique used to measure D/H is not currently limited by systematic
uncertainties, there may be other effects that could potentially bias the
determination of the primordial D/H abundance. For example, the work
by \citet{Coo14} is based on just 5 systems where the D/H abundance
can be measured with high precision, and this sample must be expanded
in the future to overcome the effects of small number statistics. Furthermore,
it is necessary to explore potential astrophysical uncertainties that could
systematically bias the results or contribute to the sample dispersion.
Perhaps the two dominant astrophysical uncertainties that might bias
a deuterium abundance measurement are:
(1) The differential ionization potential of deuterium and hydrogen,
${\rm D}_{\rm IP}-{\rm H}_{\rm IP}\simeq0.0037\,{\rm eV}$,
introduces a small systematic bias under the assumption
that D/H$\equiv$\DI/\HI.
(2) The astration of deuterium during the chemical evolution of
galaxies, can systematically lower a galaxy's D/H ratio.

In this paper, we assess the deuterium ionization correction, that may
introduce a bias in the determination of the primordial D/H ratio. The only
published investigation of the deuterium ionization correction in the context
of the primordial element abundances was conducted by \citet{Sav02},
and the issue does not seem to have been investigated further since
that work. Under the simplifying assumption of ionization balance
for deuterium in gas at $\sim10^{4}$\,K, \citet{Sav02} concluded that
D/H$\,=\,$\DI\,/\,\HI\ in mostly neutral regions (e.g. DLAs with a high
\HI\ column density), and D/H$\,\approx\,$0.996\,\DI\,/\,\HI\ in mostly
ionized regions (e.g. sub-DLAs). The correction for sub-DLAs could
therefore be as large as 0.002~dex, which is $\sim1/3$ of the
current measurement precision. Given that the current D/H measurement
precision is nearing the magnitude of the ionization correction, a more
detailed investigation into this potential bias over a much larger range
in parameter space is warranted.

In this paper, we present the results from a suite of calculations to determine
the D/H ionization correction as a function of the \HI\ column density and level
of ionization. In Section~\ref{sec:model}, we outline the details of our
photoionization simulations and compare our code to \textit{Cloudy}.
The results of our calculations are presented in Section~\ref{sec:results},
where we provide fitting formulae to determine the deuterium ionization
correction using observable quantities. We discuss our findings in
Section~\ref{sec:disc}, before summarising our main conclusions
in Section~\ref{sec:conc}.

Throughout this paper, we adopt a flat, $\Lambda$ cold dark matter cosmology,
with parameters estimated by the \citet{Efs15} analysis of the cosmic microwave
background temperature fluctuations. Specifically, we use the parameters
$h=0.673$, $\Omega_{\rm B}=0.0491$, and $\Omega_{\rm M}=0.315$,
which are now known to within $\sim1$ per cent.

\section{Photoionization Simulations}
\label{sec:model}

To calculate the relative ionization of deuterium and hydrogen,
we have developed a software package that provides an
approximate model of the gas distribution and ionization of
a metal-poor DLA. Our relatively simple calculations are quantitively
similar to the \textit{Cloudy} photoionization software \citep{Fer13}.
The necessary improvements that our code offers over \textit{Cloudy}
include: (1) The atomic physics of the deuterium atom, which are not currently
included in \textit{Cloudy}; and (2) We model the gas distribution
in hydrostatic equilibrium with a putative dark matter halo. We describe
the details of our model calculations in the following subsections.

\subsection{Halo Model}

We consider gas that is embedded within a spherically symmetric
Navarro-Frenk-White (NFW; \citealt{NFW96}) dark matter halo with
a radial density profile given by:
\begin{equation}
\rho_{\rm d}(x) = \frac{\rho_{\rm ds}}{x\,(1+x^2)},\qquad~x=r/r_{\rm s}
\end{equation}
where $\rho_{\rm ds}$ and $r_{\rm s}$ correspond to the characteristic
scale density and scale radius of the dark matter halo, respectively.
The mass of dark matter enclosed within radius, $r$, is equal to
$M_{\rm d}(<~\!\!r)~=~3 M_{\rm ds} f_{\rm M}(x)$, where:
\begin{equation}
f_{\rm M}(x) = {\rm ln}(1+x) - \frac{x}{1+x},
\end{equation}
and $M_{\rm ds}=4\pi\rho_{\rm ds}r_{s}^{3}/3$.
Throughout this paper, we refer to the virial radius of a halo ($r_{200}$) as the
radius where the average dark matter density is $200$ times the critical density
of the Universe ($\rho_{\rm crit}$), and is related to the virial mass of the halo by
the expression
\begin{equation}\label{eqn:m200}
M_{200} = \frac{4\pi r_{200}^{3}}{3}200\rho_{\rm crit}
\end{equation}
Therefore, for a given virial mass, $M_{200}$, we calculate $r_{200}$
using Eq.~\ref{eqn:m200}, and use the mass-concentration relation
provided by \citet{Pra12} to estimate the halo concentration parameter,
$c_{200}~=~r_{200}/r_{\rm s}$, and hence determine the scale
radius of the halo.

\subsection{Gas Distribution}

We model the gas density profile, $\rho_{\rm g}(r)$, in hydrostatic
equilibrium with a potential $\varphi(r)$, such that $dP(r)=-\rho_{\rm g}(r) {\rm d}\varphi(r)$.
We assume that the pressure profile of the gas comprises a thermal and a turbulent
component, such that
\begin{eqnarray}
\label{eqn:presdens}
P(r) &=& P_{\rm ther}(r)+P_{\rm turb}(r)\nonumber\\
P(r) &=& \frac{k_{\rm B}\rho_{\rm g}(r)T(r)}{m_{\rm H}\mu(r)}+\frac{3\rho_{\rm g}(r)b_{\rm turb}^{2}}{4}
\end{eqnarray}
where $T(r)$ is the radial temperature profile, $k_{\rm B}$ is the Boltzmann constant,
$m_{\rm H}$ is the proton mass, $\mu(r)$ is the mass per particle
(which has a radial dependence due to the ionization state of the gas), and
$b_{\rm turb}$ is the Doppler parameter of the gas. For this study, we use a
typical Doppler parameter of $b_{\rm turb}=3\,{\rm km~s}^{-1}$, as measured
recently for a sample of low metallicity DLAs \citep{Coo15}. These authors
also found that turbulent pressure is subdominant relative to thermal pressure
for the metal-poor DLAs in their study ($P_{\rm turb}/P_{\rm th}\sim0.1$).
Thus, a 10 per cent change in the adopted value of $b_{\rm turb}$ changes the
total pressure by $\sim 1$ per cent. Our conclusions are therefore insensitive
to the choice of $b_{\rm turb}$.

Substituting the pressure profile into the equation for hydrostatic equilibrium yields
\begin{equation}
\label{eqn:hse}
{\rm d}P/P = -{\rm d}\varphi/u_{\rm g}^2,
\end{equation}
where
\begin{equation}
u_{\rm g}^{2} = \frac{k_{\rm B}T(r)}{m_{\rm H}\mu(r)}+\frac{3b_{\rm turb}^{2}}{4}
\end{equation}
Under the simplifying assumption that the gas self-gravity
does not affect the density distribution and hence the ionization
structure of the gas\footnote{Including the self-gravity of the gas is computationally
demanding, and we cannot explore these effects in this paper. Having said that, in
Section~\ref{sec:disc} we show that the D/H ionization correction is insensitive
to the gas distribution. Thus, neglecting the gas self-gravity does not affect
our conclusions.},
${\rm d}\varphi=u_{\rm s}^2f_{\rm M}(x){\rm d}x/x^2$, where
$u_{\rm s}^{2}=G\,M_{\rm ds}/r_{s}^{2}$.
Integrating Eq.~\ref{eqn:hse} therefore yields a simple
equation for the pressure profile of the gas embedded
in a dark matter halo,
\begin{equation}
\label{eqn:presprof}
P(x) = P_{0}\,\exp\bigg(-\int_{0}^{x} \!\frac{u_{\rm s}^{2}}{u_{\rm g}^{2}(x)}\frac{f_{\rm M}(x)}{x^2}{\rm d}x\bigg)
\end{equation}
where $P_{0}$ is the central gas pressure.

Within $r_{200}$, we assume that each halo contains a gas mass
$M_{\rm g} = f_{200}\,M_{200}\,\Omega_{\rm B}/(\Omega_{\rm M}-\Omega_{\rm B})$
where $\Omega_{\rm M}$ and $\Omega_{\rm B}$ are the universal
density of matter and baryons respectively \citep{Coo14,Efs15}.
We use the scaling constant, $f_{200}<1$, to explore models where
a dark matter halo contains fewer baryons than the universal baryon
fraction. Once $M_{\rm g}$ is specified, the gas density is normalised
such that,
\begin{equation}
\label{eqn:gasmass}
M_{\rm g} = 4\pi r_{\rm s}^{3} \int_{0}^{c_{200}} \!\rho_{\rm g}(R)\,R^2{\rm d}R
\end{equation}

\subsection{Ionization Balance}

In this paper, we consider models that may represent the most metal-poor DLAs currently known.
Given the presumably minimal level of recent star formation in these systems, it
is reasonable to assume that the local sources of ionizing photons are subdominant
relative to the extragalactic background. In what follows, we have
therefore assumed that the surface of the most metal-poor DLAs is illuminated
solely by the \citet{HarMad12} ultraviolet/X-ray background radiation from quasars
and galaxies with intensity $J_{0}(\nu)$.
We also explore simple power-law models for the ionizing background, which
are equivalent to the \texttt{`table power law slope $\alpha$'} command within
\textit{Cloudy}, where $f(\nu)\propto\nu^{\alpha}$.

\begin{figure}
  \centering
 {\includegraphics[angle=0,width=80mm]{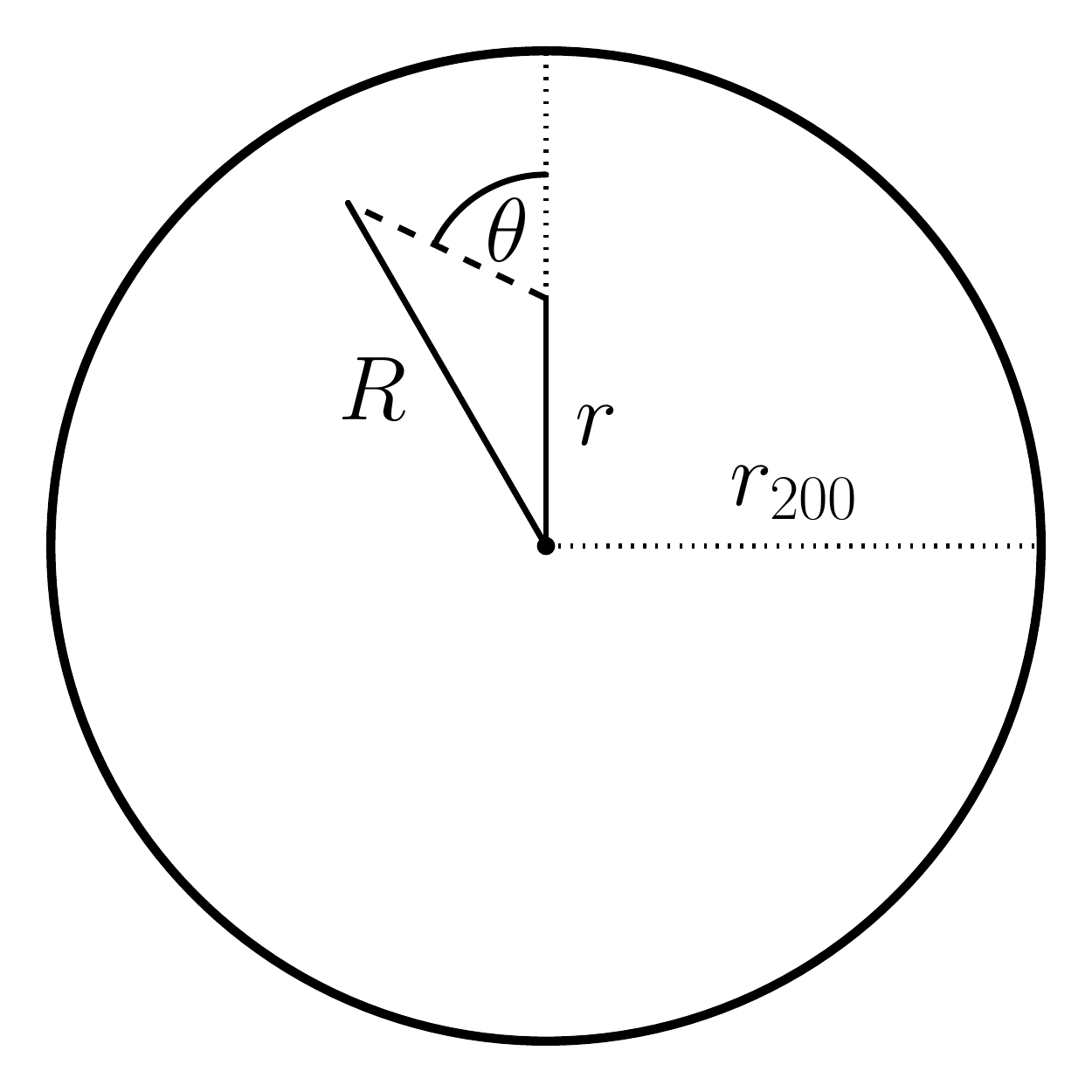}}\\
  \caption{
The coordinate system used to calculate the angular dependence
of the column density of each species at a given radial coordinate
(see Eq.~\ref{eqn:coldensa} and Eq.~\ref{eqn:coldensb}).
  }
  \label{fig:geom}
\end{figure}

We assume that the distributions of gas and dark matter are
spherically symmetric. Under this assumption, gas at a
distance $r$ from the centre of a dark matter halo is irradiated
by background photons from all directions. The radiation field
is therefore attenuated by a different amount depending on the
density distribution of the gas along a given direction. If $J_{0}(\nu)$
represents the unattenuated radiation field at a frequency $\nu$,
and $2\pi\sin(\theta){\rm d}\theta/4\pi$ is the fraction of sky with an
optical depth $\tau(\nu,r,\theta)$, then the intensity of the radiation
field at a distance $r$ from the centre of the cloud is given by
\begin{eqnarray}\label{eqn:jnur}
J(\nu,r) &=& J_{0}(\nu) \int_{0}^{\pi} \!\frac{2\pi\sin(\theta)}{4\pi}\exp[-\tau(\nu,r,\theta)]{\rm d}\theta\\
             &=& \frac{1}{2}\,J_{0}(\nu) \int_{-1}^{+1} \!\exp[-\tau(\nu,r,\mu)]{\rm d}\mu\\
\tau(\nu,r,\mu)&=&\sum\limits_{i} \sigma_{i}(\nu)N_{i}(r,\mu)\label{eqn:jnurn}
\end{eqnarray}
where $\mu=\cos\theta$, and $i$ corresponds to every element/ion stage that
we consider in this work. Unfortunately, our calculations are computationally too
demanding to include every ionization stage for all elements with atomic number
$<30$ (unlike \textit{Cloudy}). We have therefore incorporated only a
handful of the most abundant metals that are commonly observed in metal-poor
DLAs, including \HI, \DI, \HeI, \HeII, \CI--\CIV, \NI--\NIV, \OI--\OIV, \MgI--\MgIV, \SiI--\SiIV.
We have adopted the photoionization cross-section, $\sigma_{i}(\nu)$, of each element
using the compilation by \citet{Ver96}.

At a distance $r$ from the centre of the halo, the column density of
species $i$ in the direction $\theta$ can be calculated by integrating
the volume density of species $i$, $n_{i}(R)$, from a radial distance $r$ to the
virial radius\footnote{We assume that the gas beyond $r_{200}$ is
transparent to ionizing photons.} (i.e. integrate along the path defined
by the dashed line in Fig.~\ref{fig:geom}):
\begin{eqnarray}
\label{eqn:coldensa}
N_{i}(r,\mu)&=&\int\limits_{r}^{r_{200}}\!\frac{R\,n_{i}(R)\,{\rm d}R}{\sqrt{R^2-r^2(1-\mu^2)}}\\
\label{eqn:coldensb}
N_{i}(r,\mu)&=&\!\!\!\!\!\!\!\int\limits_{r\sqrt{1-\mu^2}}^{r}\!\!\!\!\!\frac{R\,n_{i}(R)\,{\rm d}R}{\sqrt{R^2-r^2(1-\mu^2)}}+
\!\!\!\!\!\!\!\int\limits_{r\sqrt{1-\mu^2}}^{r_{200}}\!\!\!\!\!\frac{R\,n_{i}(R)\,{\rm d}R}{\sqrt{R^2-r^2(1-\mu^2)}}
\end{eqnarray}
where Eq.~\ref{eqn:coldensa} is used for $0\leq\theta\leq\pi/2$
and Eq.~\ref{eqn:coldensb} is used for $\pi/2<\theta\leq\pi$.
The rate of primary ionizing photons at a given radius from the
cloud centre of each species is then
\begin{equation}
\label{eqn:gasmass}
\Gamma_{\rm p}(r) = 4\pi \int_{\nu_{i}}^{\infty} \!\frac{J(\nu,r)}{h\nu}\sigma_{i}(\nu){\rm d}\nu
\end{equation}
where $h\nu_{i}$ is the ionization energy\footnote{We use the ionization energies
provided by the National Institute of Standards and Technology (NIST):\\\texttt{http://physics.nist.gov/PhysRefData/ASD/ionEnergy.html}} of ion $i$. 
Each radiation field considered in this work is finely interpolated around the ionization energy
of each ion to ensure numerical accuracy in the integrations.

The rate of collisional ionization, $\Gamma_{\rm ci}(T)$, is incorporated in our models
using the \citet{Der07} rate calculations. Photoionization of
\HI, \HeI, and \HeII\ from recombinations of these species, $\Gamma_{\rm r}$,
is included using equations B1, B2, B3, B6, and B7 from \citet{Jen13}.
We have also included the contribution from secondary collisional
ionizations from energetic primary photoelectrons, $\Gamma_{\rm s}$,
using the prescription outlined in \citet[][cf. \citealt{ShuVan85}]{RicGneShu02}.
The sum of these four ionization rates for each element and ion stage is denoted
$\Gamma({\rm X}^{\rm q+})$ at each radial position.
Radiative and dielectronic recombination rates, $\alpha_{\rm r}(T)$, are
calculated using the method outlined by \citet{Bad03} and
\citet{Bad06}\footnote{See the following website for further details:\\\texttt{http://amdpp.phys.strath.ac.uk/tamoc/DATA/}}.
Finally, rates for charge exchange ionization and recombination are determined using the \citet{KinFer96}
database\footnote{With updates available from the following website:\\\texttt{http://www-cfadc.phy.ornl.gov/astro/ps/data/}}.
We define the charge transfer coefficients for reactions of ion ${\rm X}^{\rm q+}$ with H or He using the characters $C$ or $D$
respectively, according to the following definitions:
\begin{eqnarray}
C_{\rm i}({\rm X}^{\rm q+},T) &=& {\rm X}^{\rm q+}+{\rm H}^{+}\to{\rm X}^{\rm (q+1)+}+{\rm H}^{0}\\
C_{\rm r}({\rm X}^{\rm q+},T) &=& {\rm X}^{\rm q+}+{\rm H}^{0}\to{\rm X}^{\rm (q-1)+}+{\rm H}^{+}\\
D_{\rm i}({\rm X}^{\rm q+},T) &=& {\rm X}^{\rm q+}+{\rm He}^{+}\to{\rm X}^{\rm (q+1)+}+{\rm He}^{0}\\
D_{\rm r}({\rm X}^{\rm q+},T) &=& {\rm X}^{\rm q+}+{\rm He}^{0}\to{\rm X}^{\rm (q-1)+}+{\rm He}^{+}
\end{eqnarray}

The rates for charge exchange between deuterium and hydrogen were derived from the data listed in Table~1 of \citet{Sav02}.
As we discuss in Section~\ref{sec:disc}, the \emph{relative} reaction rate for deuterium charge exchange
ionization and recombination largely determines the deuterium ionization correction.
Since the ionization potential for H is less than D, the rate of the
endothermic reaction $C_{\rm i}({\rm D}^{\rm 0},T)$ should always
be less than that of the exothermic reaction $C_{\rm r}({\rm D}^{\rm +},T)$.
We note that the approximate fitting formulae
provided by \citet{Sav02} reverse the endothermic and exothermic
nature of the reaction in the temperature range 4,500\,K -- 100,000\,K
(i.e. $C_{\rm i}({\rm D}^{\rm 0},T)$ is larger than $C_{\rm r}({\rm D}^{\rm +},T)$).
This approximation leads to a notable and incorrect change to both the magnitude
and sign of the deuterium ionization correction.
We therefore adopt their recommended fitting function for deuterium
charge exchange ionization:
\begin{equation}\label{eqn:dcti}
C_{\rm i}({\rm D}^{\rm 0},T) = 2\times10^{-10}\,T^{0.402}\,\exp\,(-37.1/T)\,-\,3.31\times10^{-17}\,T^{1.48}
\end{equation}
and adopt the following form for deuterium charge exchange recombination,
under the assumption of chemical equilibrium:
\begin{equation}\label{eqn:dctr}
C_{\rm r}({\rm D}^{\rm +},T) = C_{\rm i}({\rm D}^{\rm 0},T) \times \exp\,(42.915/T)
\end{equation}
where $42.915\,{\rm K}$ is the difference in ionization potential of D relative to H.
Eq.~\ref{eqn:dctr} provides an accurate description of the relative
reaction rate data (to within 0.6~per~cent) compiled by \citet{Sav02} in the
temperature range 2\,K -- 200,000\,K, and ensures that
$C_{\rm i}({\rm D}^{\rm 0},T)$ is always less than $C_{\rm r}({\rm D}^{\rm +},T)$.

Throughout this work, we assume that each element is in
ionization equilibrium, such that:
\begin{eqnarray}
\label{eqn:ionbalance}
\big[ \Gamma({\rm X}^{\rm q+}) + C_{\rm i}({\rm X}^{\rm q+},T)\,n({\rm H}^{+}) + D_{\rm i}({\rm X}^{\rm q+},T)\,n({\rm He}^{+})\big]\,n({\rm X}^{\rm q+}) = \nonumber \\
\big[ \alpha_{\rm r}({\rm X}^{\rm q+},T)\,n_{\rm e} + C_{\rm r}({\rm X}^{\rm (q+1)+},T)\,n({\rm H}^{0}) + D_{\rm r}({\rm X}^{\rm (q+1)+},T)\,n({\rm He}^{0})\big]\times\nonumber\\
\,n({\rm X}^{\rm (q+1)+})
\end{eqnarray}
for the q ionization states of a given element X. The electron density, $n_{\rm e}$,
is calculated from H and He ionizations, and we ignore the contribution of electrons
from the ionization of metals\footnote{Electrons from metals offer a negligible contribution
at the extremely low metallicities, and predominantly neutral gas that we consider in this work.}.
Considering all ionization states for a given element, Eq.~\ref{eqn:ionbalance} represents
a set of simultaneous equations that can be solved for the fractional ionization of
a given species at each radial coordinate, ${\cal F_{\rm X^{q+}}}\equiv n({\rm X^{q+}})/n({\rm X})$.

\begin{figure*}
  \centering
 {\includegraphics[angle=0,width=160mm]{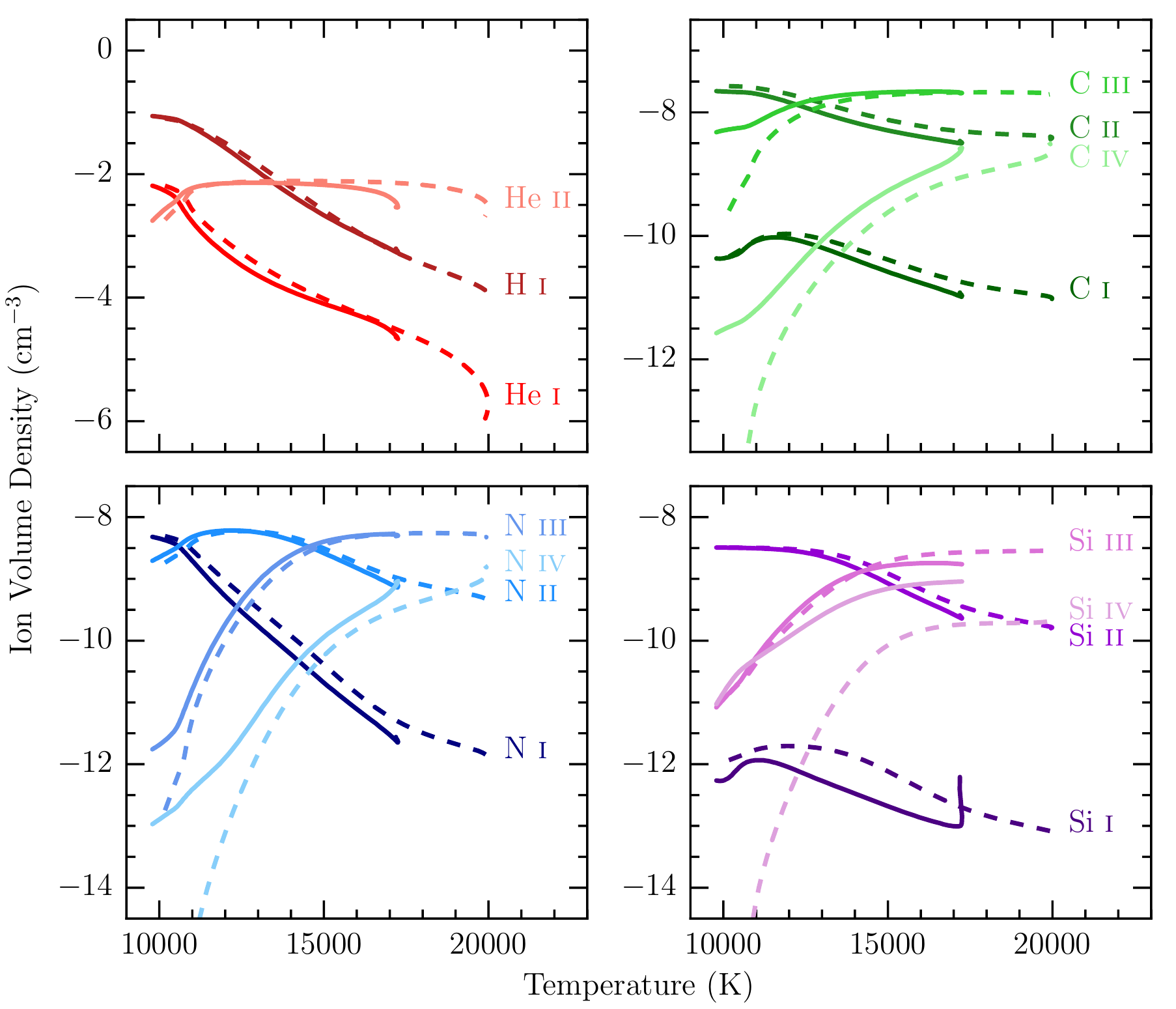}}\\
  \caption{
The temperature-density relation for a plane parallel slab of
constant density gas. The dashed lines indicate the results from
our model calculations, while the solid lines were calculated
using the \textit{Cloudy} photoionization software. In general,
there is a good agreement between the two codes.
  }
  \label{fig:cloudy}
\end{figure*}

\subsection{Thermal Equilibrium}

In addition to ionization equilibrium, we assume that the gas is in thermal equilibrium
such that the total heating rate exactly balances the total cooling rate at each radial coordinate.
The heating rate includes contributions from both photoionization heating and secondary
heating by primary photoelectrons. The photoionization heating rate for each chemical
element at a given radius is given by:
\begin{equation}
{\cal H}({\rm X}^{\rm q+}) = n({\rm X}^{\rm q+})\int\limits_{\nu_{\rm i}}^{\infty}\!\sigma_{\rm i}(\nu)~J(\nu,r)~\frac{h\nu-h\nu_{\rm i}}{h\nu}~{\rm d}\nu
\end{equation}
and the total heating rate is summed over all species. The cooling rate
includes contributions from collisional excitation/ionization
cooling, single electron and dielectronic recombination cooling,
Bremsstrahlung cooling and Compton heating/cooling
(see equations 12a-17 from \citet{Cen92} for a complete list of the adopted cooling formulae).
Given the very low metallicity of the gas being probed (typically $<1/100$ solar), we have
ignored metal cooling in our calculation, and do not expect this to considerably alter
our conclusions.

\subsection{Numerical Method}
\label{sec:nummeth}
Our calculations are qualitatively similar to those presented by \citet{Kep97} and \citet{Ste02}.
We use an iterative procedure to solve the equations of thermal and ionization equilibrium.
Each model calculation is initialised with a primordial $^{4}$He mass fraction
${\rm Y}_{\rm P}=0.25$, and a primordial deuterium abundance
$\log\,({\rm D/H})_{\rm P}=-4.60$. We assume that the model DLAs have a
metal abundance distribution that is consistent with the solar abundance
pattern, scaled to a metallicity $1/1000$ solar\footnote{The DLAs that are
typically used to determine the primordial D/H abundance have a metallicity
$<1/100$ solar. Since at metallicities less than 1/100 solar
the metals are trace components that do not
contribute significantly to the thermal properties of the gas,
the exact value of the adopted metallicity is
not expected to affect our conclusions.}. We then select a dark
matter halo mass ($M_{200}$), and a scaling factor ($f_{200}$) to determine
the total mass of baryons within the halo virial radius. The initial gas temperature
is set to 20,000\,K, and we assume that the gas is mostly ionized.
As we discuss in
Section~\ref{sec:disc}, the most important factor in determining the magnitude of
the D/H ionization correction is D$\leftrightarrow$H charge exchange, which does
not depend on any of the above assumptions.

These initial parameter values allow us to numerically solve the integral
in Eq.~\ref{eqn:presprof}, and determine the pressure profile of the gas.
Using Eq.~\ref{eqn:presdens}, we can then calculate the gas density profile.
The gas density profile establishes the radiation field at a given
radial coordinate (Eq.~\ref{eqn:jnur}-\ref{eqn:coldensb}). We then solve for ionization equilibrium
to calculate the fractional ionization of each species, ${\cal F_{\rm X^{q+}}}$,
at all radial coordinates. To increase the efficiency of our computation, we
sub-iterate over the solution to the equations of ionization balance, holding the
temperature and radiation field constant at each coordinate, until the
relative difference of ${\cal F_{\rm X^{q+}}}$ between each sub-iteration
for all species is less than $10^{-5}$.

The photoheating rate at each radial position is calculated using the values
of $n({\rm X}^{\rm q+})$ determined from ${\cal F_{\rm X^{q+}}}$ and the
corresponding radiation field at this position. We then estimate the temperature
profile of the gas assuming thermal equilibrium. With this new temperature
profile, we recalculate the integral in Eq.~\ref{eqn:presprof} for the pressure
profile. We then iterate the procedure described above until the relative
difference of ${\cal F_{\rm X^{q+}}}$ between successive iterations for all
species is less than $10^{-5}$.

\begin{figure*}
  \centering
 {\includegraphics[angle=0,width=160mm]{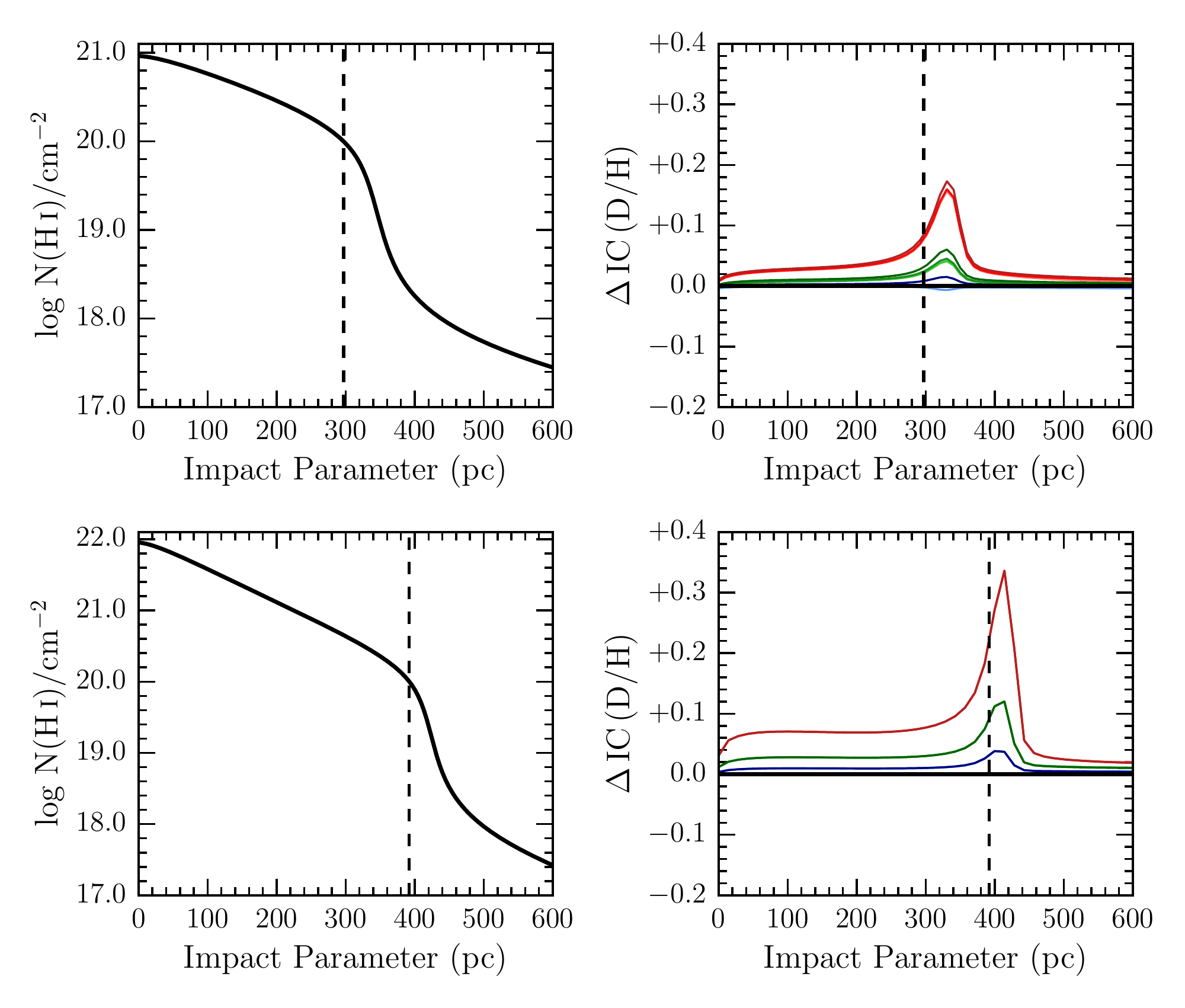}}\\
  \caption{
We performed a convergence test on two model halos to optimise
the computational speed and accuracy of our calculations.
The results presented in the top panels correspond to a
$10^{8.4}~{\rm M}_{\odot}$ dark matter halo with a baryon fraction
$f_{200}=1.0$ (i.e. the Universal baryon fraction), and the bottom panels
present the results for a $10^{8.9}~{\rm M}_{\odot}$ dark matter halo with
$f_{200}=0.3$. Both calculations were performed at redshift $z=3$, with
an isotropic, \citet{HarMad12} incident radiation field. The solid black lines
in the left panels display the \HI\ column density profile of the halo.
In the right panels, we illustrate the \emph{fractional} change of the deuterium
ionization correction between a high resolution
calculation (solid black line), and various lower resolution calculations
(see Eq.~\ref{eqn:dhfracconv} and text for further details); the red, green,
and blue curves correspond
to a radial sampling, $n_{\rm r}$ = 500, 1000, and 2000 respectively.
Light to dark shades of each coloured curve
(almost indistinguishable over the plotted radial range)
correspond to an angular
sampling $n_{\mu}$ = 30, 90, and 180, respectively.
The vertical dashed line in all panels corresponds to the radius where
the \HI\ column density $N({\rm H\,\textsc{i}})=10^{20}~{\rm cm}^{-2}$.
Note that $\Delta$IC(D/H) represents the \emph{fractional} change
in the deuterium ionization correction (i.e. $+0.1\equiv10$~per~cent
uncertainty in the value of the correction; see Eq.~\ref{eqn:dhfracconv}).
  }
  \label{fig:converge}
\end{figure*}

\subsection{Model Observables}

In general, information on metal-poor DLAs is restricted to a
single line-of-sight to a background quasar. For a quasar
that intersects a metal-poor DLA at an impact parameter $b$,
with respect to the centre of the dark matter halo,
the measured column density of the species X$^{\rm q+}$ is
given by
\begin{equation}
\label{eqn:coldensip}
N_{\rm X^{\rm q+}}(b) = 2\int\limits_{b}^{\infty}\!\frac{r}{\sqrt{r^2-b^2}}\,n({\rm X^{\rm q+}},r)\,{\rm d}r
\end{equation}

\subsection{Numerical Stability and Convergence}

\subsubsection{Stability}

We performed a series of checks to ensure the reliability of our code.
We first tested the implementation of our ionization and thermal equilibrium
equations by modelling a plane parallel slab of constant density gas
($n({\rm H})=0.1~{\rm cm}^{-3}$) illuminated on one side by a
power law radiation field ($f(\nu)\propto\nu^{\alpha}$) with
index $\alpha=-1$. Our simulation was stopped once a neutral
H column density of $10^{20.3}$ \HI\ atoms cm$^{-2}$ was reached.
These parameters correspond to a typical DLA, irradiated on one
side by a quasar. The relationships between temperature and density
for each chemical species in our calculations are shown by the
dashed curves in Fig.~\ref{fig:cloudy}.
We have performed a nearly identical calculation with the \textit{Cloudy}
photoionization software \citep{Fer13}. The results from the \textit{Cloudy}
simulation are shown as solid lines in Fig.~\ref{fig:cloudy}.

There is a good agreement between the two codes for all ion stages of the most
abundant elements, including H, He, C, and N (as well as O, not shown). The
primary disagreement between \textit{Cloudy} and our code for these elements
is seen for the high ionization stages at low temperatures. This difference is due
to the large number of physical processes that are included in
\textit{Cloudy}, but are not present in our code. The most important
temperature regime to correctly model corresponds to where the
ion density is most abundant, since this regime tends to be the dominant
contribution to the column density. For these elements, our code produces
acceptable results. The temperature-density relationships for the
Si ions (as well as Mg, not shown), on the other hand, show a higher level of disagreement
between the two codes, particularly for the higher ion stages. This
discrepancy is likely due to the exclusion of elements with a
similar abundance to Si (e.g. Ne, Fe), as well as the exclusion
of several physical processes included in \textit{Cloudy} (discussed above)
and small differences between the heating and cooling functions used
by the two codes. We note, however, that there is an acceptable
agreement between the two codes for the ions of interest to our
study (including \HI, \CII, \CIII, \NI, \NII, \SiII\ and \SiIII).
Specifically, the difference between our code and \textit{Cloudy}
for the column density ratio of successive ion stages are:
\begin{eqnarray}
\!\!\!\!\!\!\!\!\!\!\!\!\!&&\log~N(\mathrm{C\,\textsc{iii})/N(C\,\textsc{ii}})_{\rm our~code}~-~\log~N(\mathrm{C\,\textsc{iii})/N(C\,\textsc{ii}})_{\rm cloudy}=-0.09\nonumber\\
\!\!\!\!\!\!\!\!\!\!\!\!\!&&\log~N(\mathrm{N\,\textsc{ii})/N(N\,\textsc{i}})_{\rm our~code}~-~\log~N(\mathrm{N\,\textsc{ii})/N(N\,\textsc{i}})_{\rm cloudy}=+0.02\nonumber\\
\!\!\!\!\!\!\!\!\!\!\!\!\!&&\log~N(\mathrm{Si\,\textsc{iii})/N(Si\,\textsc{ii}})_{\rm our~code}~-~\log~N(\mathrm{Si\,\textsc{iii})/N(Si\,\textsc{ii}})_{\rm cloudy}=+0.40\nonumber
\end{eqnarray}
Therefore, as discussed above, we conclude that the Si ratio shows
the highest level of disagreement between the two codes\footnote{We
later show in Section~\ref{sec:results}, that this level of disagreement
for Si changes the D/H ionization correction by $<0.0005$~dex.}.

\subsubsection{Convergence}
\label{sec:converge}

We then performed a convergence and optimisation study, to determine
the number of radial coordinates needed to accurately evaluate the
numerical integrations over radius and $\cos\theta$ for the NFW geometry. We conducted two
simulations: (1) A $10^{8.4}~{\rm M}_{\odot}$ dark matter halo with $f_{200}=1.0$ (top panels of Fig.~\ref{fig:converge}); and
(2) a $10^{8.9}~{\rm M}_{\odot}$ dark matter halo with $f_{200}=0.3$ (bottom panels of Fig.~\ref{fig:converge}).
Both calculations assume that the halo is exposed to an isotropic \citet{HarMad12} background
radiation field at redshift $z=3$.
The former and latter halos produce a maximum neutral hydrogen column density of
$N({\rm H\,\textsc{i}})\simeq10^{21}~{\rm cm}^{-2}$ and
$N({\rm H\,\textsc{i}})\simeq10^{22}~{\rm cm}^{-2}$ respectively, for a line-of-sight that passes
directly through the centre of the halo (corresponding to $b=0$ in Eq.~\ref{eqn:coldensip}).
The \HI\ column density profiles as a function of impact parameter for these two halos are
shown in the left panels of Fig.~\ref{fig:converge}, where the vertical dashed line corresponds
to the radius where $N({\rm H\,\textsc{i}})=10^{20}~{\rm cm}^{-2}$.

We began the simulations with a very fine sampling of $n_{\mu}=720$ angular coordinates over $\cos\theta$,
and $n_{\rm r}=5000$ radial coordinates linearly spaced between the centre of the halo and $r_{200}$.
We then explored various combinations of these samplings to optimise the efficiency of our calculations whilst
maintaining numerical accuracy. Our grid spanned $n_{\mu}~=~[30, 90, 180]$ and $n_{\rm r}~=~[500, 1000, 2000]$.
The right panels of Fig.~\ref{fig:converge} illustrate the results of our convergence study, where
$\Delta{\rm IC(D/H})$ is the \emph{fractional} change to the ionization correction for different
choices of the numerical integration parameters:
\begin{equation}
\label{eqn:dhfracconv}
\Delta{\rm IC(D/H})\equiv\frac{{\rm IC(D/H})_{\rm 720,5000}-{\rm IC(D/H})_{n_{\mu},n_{\rm r}}}{{\rm IC(D/H})_{720,5000}}
\end{equation}
where IC(D/H) is defined below, in Eq.~\ref{eqn:icdh}.
As expected, the choice of $n_{\rm r}$ is most sensitive to the region where the neutral gas is
becoming more ionized, somewhat below the classical DLA threshold of
$N({\rm H\,\textsc{i}})=10^{20.3}~{\rm cm}^{-2}$. On the other hand, the choice of $n_{\rm \mu}$
only becomes important when $r~\gg~r_{\rm DLA}$, well beyond the radial regime of interest
to this study.
The optimal combination for our study is therefore $n_{\mu}=30$ angular coordinates and
$n_{\rm r}=1000$ radial coordinates, which provided a deuterium ionization correction that
is accurate to within $\sim10$ per cent when $N({\rm H\,\textsc{i}})\ge10^{20}~{\rm cm}^{-2}$.

\section{D/H Ionization Correction}
\label{sec:results}

We now discuss the results from our calculations to determine the deuterium ionization correction.
In what follows, we define the true value of the D/H abundance to be
\begin{equation}\label{eqn:icdh}
\log_{10} {\rm D/H} = \log_{10} N({\rm D}\,\textrm{{\sc i}})/N({\rm H}\,\textrm{{\sc i}}) + {\rm IC(D/H)}
\end{equation}
where ${\rm IC(D/H)}$ is the deuterium ionization correction derived from our model calculations.
We adopt a typical value for the `true' deuterium abundance, $\log_{10} {\rm D/H}=-4.60$. We have
computed a series of calculations with a simple plane parallel geometry as well as a more
`realistic' geometry where the gas is confined to an NFW halo.

\subsection{Plane Parallel Models}
\label{sec:ppmod}

The simplest geometry to consider is a uniform, plane parallel,
constant volume density slab of gas, illuminated on one side. For
our calculations, we have assumed the gas slab is irradiated
by the \citet{HarMad12} radiation field at redshift $z=3$, incident
normal to the surface of the slab. We computed a grid of calculations
covering a range in
H volume density ($-2.0\le\log\,n_{\rm H}/{\rm cm}^{-3}\le2.0$, in steps of 0.5 dex) and
\HI\ column density ($19.0\le\log\,N({\rm H\,\textsc{i}})/{\rm cm}^{-2}\le21.0$, in steps of 0.5 dex).
The metals
were assumed to be in solar relative proportion \citep{Asp09}, and globally scaled to a metallicity
of $1/1000~Z_{\odot}$. The depth of the slab was increased until the desired \HI\ column density
had been reached. The depth of each simulated slab was sampled linearly by 1000 values. Note that changing
the H volume density is equivalent to changing the intensity (but not the shape) of the incident radiation
field.

\begin{figure*}
  \centering
 {\includegraphics[angle=0,width=87mm]{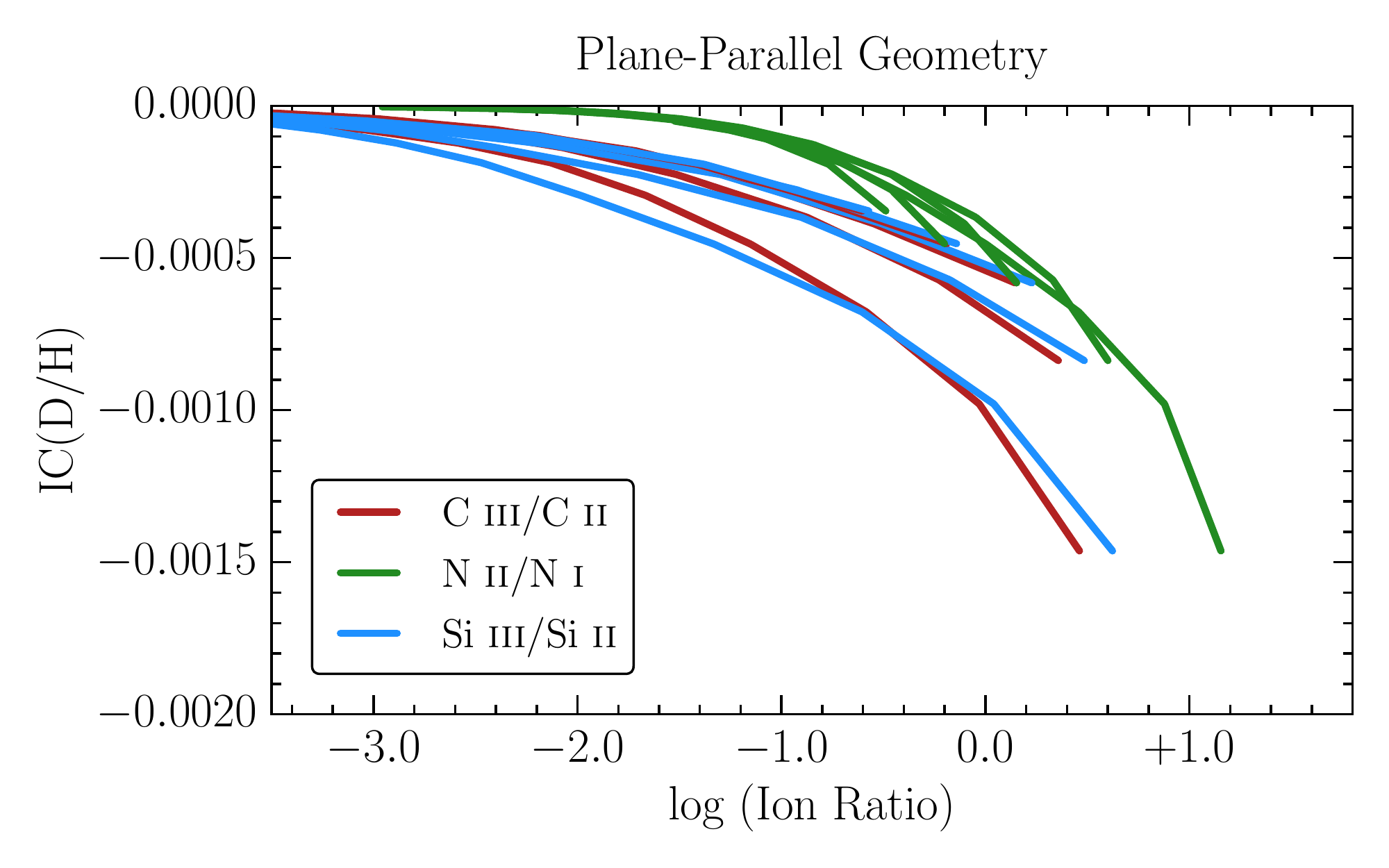}
 \includegraphics[angle=0,width=87mm]{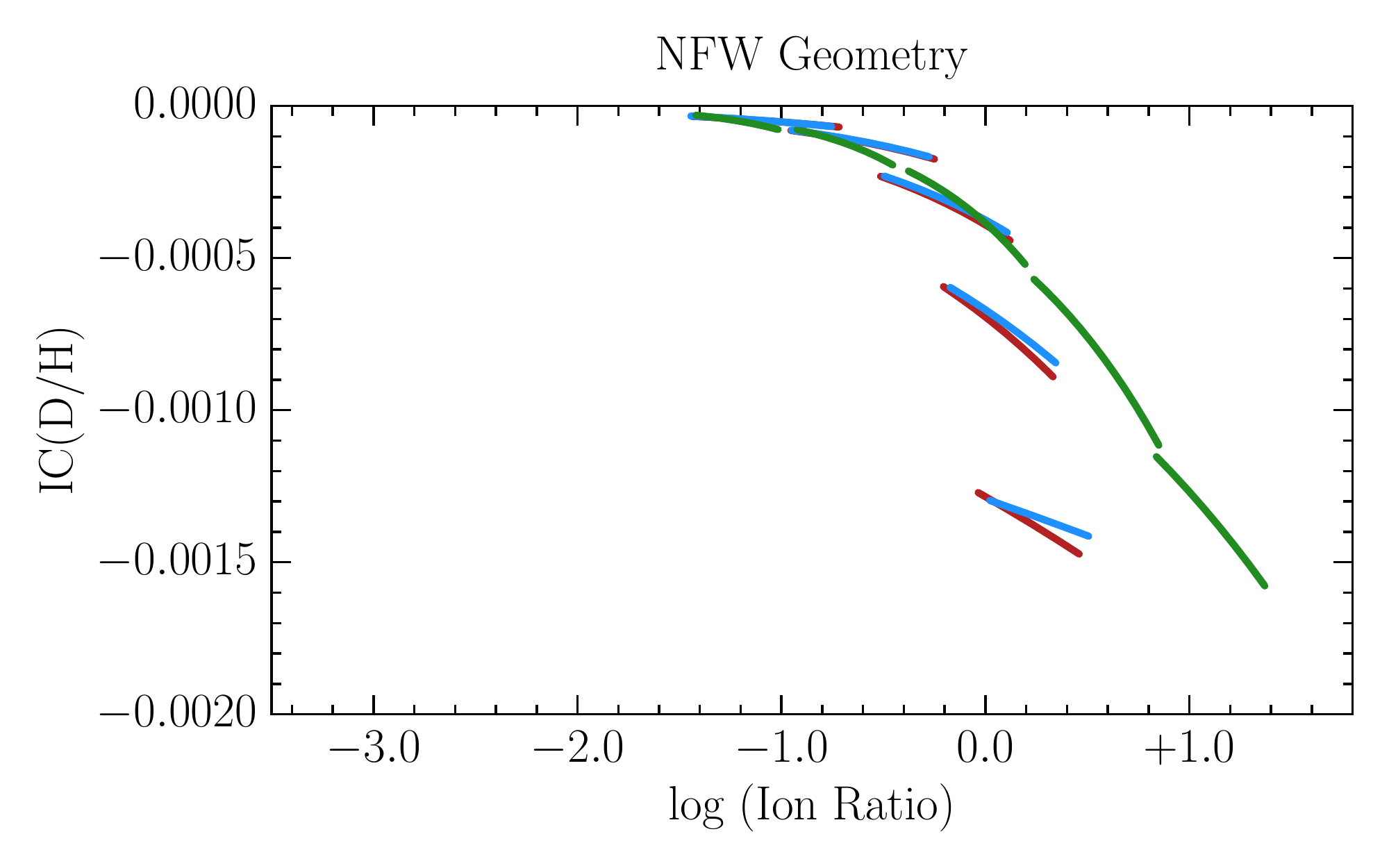}}\\
  \caption{
The deuterium ionization correction is shown for a gas cloud exposed
to the \citet{HarMad12} background radiation field. The left panel
illustrates the results for a plane parallel geometry with uniform volume
density, while the right panel illustrates the correction for a gas cloud in
hydrostatic equilibrium with an NFW potential.
The red, green and blue curves illustrate how the D/H ionization correction
depends on the \CIII/\CII, \NII/\NI, and \SiIII/\SiII\ ion ratios. Each curve
represents the correction at a given \HI\ column density. There are 5 curves
for each ion ratio, corresponding to \HI\ column densities
log~$N$(\HI)/cm$^{-2}=19.0, 19.5, 20.0, 20.5, {\rm and}\,21.0$. In the case of \NII/\NI,
the curves for different values of $N$(\HI) largely overlap with one another.
For each ion ratio, the curve with the most negative IC(D/H) values corresponds to a
log~$N$(\HI)/cm$^{-2}=19.0$; IC(D/H) becomes smaller as the \HI\ column density increases.
The curves in the left panel cover a range in H volume density
($-2.0\le\log\,n_{\rm H}/{\rm cm}^{-3}\le2.0$), while the curves in the
right panel consider a range in the physical properties of an NFW
halo (e.g. halo mass, baryon fraction, radiation intensity; see text for
further details).
  }
  \label{fig:ppic}
\end{figure*}

The results of our calculations are presented in the left panel of Fig.~\ref{fig:ppic},
where the red, green and blue curves correspond to the column density ratios
$N({\rm C\,\textsc{iii}})/N({\rm C\,\textsc{ii}})$,
$N({\rm N\,\textsc{ii}})/N({\rm N\,\textsc{i}})$, and
$N({\rm Si\,\textsc{iii}})/N({\rm Si\,\textsc{ii}})$ respectively.
In all cases, the ionization correction for deuterium is negative for the
set of H volume densities and \HI\ column densities considered in this work.
The maximum correction in our models is $-0.0015$~dex (i.e. $\sim0.4$ per cent),
corresponding to gas with a low \HI\ column density ($N$(\HI)$< 10^{20}~{\rm cm}^{-2}$)
and/or high ionization.
${\rm IC(D/H)}$ exhibits a similar dependence on the C and Si ion ratios.
The deuterium ionization correction for these ions 
depends on the values of the ion ratio and the \HI\ column density.
On the other hand, the deuterium ionization correction determined from the N ion
ratio is almost independent of the \HI\ column density. For the range of $n_{\rm H}$
and $N$(\HI) considered here, the following fitting formula
can be used to estimate the ionization correction:
\begin{equation}
\label{eqn:icpoly}
e^{{\rm IC(D/H)}} = -\sum_{m}\sum_{n}a_{nm}~(\log_{10} N(\textrm{H\,{\sc i}})/{\rm cm}^{-2} )^{n}~(\log_{10} {\rm IR})^{m}
\end{equation}
where IR corresponds to the column density ratio of successive ion stages
(for example, IR=$N({\rm C\,\textsc{iii}})/N({\rm C\,\textsc{ii}})$,
$N({\rm N\,\textsc{ii}})/N({\rm N\,\textsc{i}})$, or
$N({\rm Si\,\textsc{iii}})/N({\rm Si\,\textsc{ii}})$),
and the $a_{nm}$ coefficients are provided in
Table~\ref{tab:PPcoeff}. We caution against extrapolating these curves
beyond the appropriate ranges shown in the left panel of Fig.~\ref{fig:ppic}.

\begin{table}
\centering
\begin{minipage}[c]{0.5\textwidth}
    \caption{\textsc{plane-parallel ionization correction coefficients}}
    \begin{tabular}{lcccc}
    \hline
    \hline
    \multicolumn{5}{c}{C\,\textsc{iii}/C\,\textsc{ii}}\\
    \hline
    $n$ & $m$ & & & \\\cmidrule{2-5}
            & $0$ & $1$ & $2$ & $3$ \\
    \hline
    $0$& $114.39$ & $25.084$ & $17.063$ & $2.8237$ \\
    $1$& $-11.816$ & $-2.3171$ & $-1.6591$ & $-0.2769$ \\
    $2$& $0.2860$ & $0.05441$ & $0.03993$ & $0.00675$ \\
    \hline
    \hline
    \multicolumn{5}{c}{N\,\textsc{ii}/N\,\textsc{i}}\\
    \hline
    $n$ & $m$ & & \\\cmidrule{2-5}
        & $0$ & $1$ & $2$ & $3$ \\
    \hline
    $0$ & $144.62$ & $83.202$ & $26.766$ & $9.9952$ \\
    $1$ & $-15.688$ & $-8.8214$ & $-2.8651$ & $-0.98627$ \\
    $2$ & $0.40361$ & $0.23679$ & $0.07629$ & $0.0244$ \\
    \hline
    \hline
    \multicolumn{5}{c}{Si\,\textsc{iii}/Si\,\textsc{ii}}\\
    \hline
    $n$ & $m$ & & & \\\cmidrule{2-5}
        & $0$ & $1$ & $2$ & $3$ \\
    \hline
    $0$ & $148.77$ & $48.975$ & $17.508$ & $2.9050$ \\
    $1$ & $-15.372$ & $-4.9716$ & $-1.7676$ & $-0.2844$ \\
    $2$ & $0.3777$ & $0.1277$ & $0.04468$ & $0.00701$ \\
    \hline
    \end{tabular}
    \label{tab:PPcoeff}
\end{minipage}
\end{table}

\subsection{NFW Models}
\label{sec:nfwmod}

As shown in Section~\ref{sec:ppmod}, the D/H ionization correction
could become comparable to the measurement precision in the
near future. It is therefore necessary to test if the ionization correction
depends on the geometry of the gas. In this section, we explore a gas
geometry that might represent more closely the physical state of the gas
that gives rise to metal-poor DLAs, rather than a simple gas slab of uniform
density. In what follows, we discuss the results from modelling
metal-poor DLAs as gas in hydrostatic pressure equilibrium with
the potential of a host NFW dark matter halo.

For a metal-poor DLA that is probed by the sightline to a
background quasar, there are several unknown physical
properties that cannot be determined directly. These include:
(1) The host dark matter halo mass, $M_{200}$, of the DLA;
(2) The baryon fraction, $f_{200}$, of the halo;
(3) The intensity of the incident radiation field\footnote{Unlike the plane parallel geometry (where changes to the volume density are degenerate with changes to the intensity of the incident radiation field), the volume density in the NFW geometry is fixed by hydrostatic pressure equilibrium for a given set of input parameters. We therefore treat the intensity of the radiation field as a free parameter in the modelling procedure.}; and
(4) The shape of the radiation field.
Other properties, such as the redshift, turbulent Doppler parameter,
and metal abundances of the DLA can be inferred from observations.
We adopt typical parameter values for these quantities: $z=3$, $b_{\rm turb}=3.0~{\rm km~s}^{-1}$,
and we globally scale the metals to 1/1000~$Z_{\odot}$, in solar relative
proportions\footnote{Although metal-poor DLAs can exhibit 
deviations from solar-scaled chemical abundances
\citep[see][]{Coo11, Coo12, Coo13}, these differences
have a negligible effect on the ionization and thermal properties of the
gas, especially at low metallicity.}. To reduce the computational demand
of these calculations, we only consider the \citet{HarMad12} radiation
field shape at the appropriate redshift, and scale the intensity by a factor
of either $1/3$, $1$, or $3$. As we discuss in Section~\ref{sec:disc}, changing
the shape of the radiation field does not effect the estimated
deuterium ionization correction. We consider models that have a baryon
fraction in the range $\log_{10}~f_{200}= -1.0, -0.5$ and 0.0. Each numerical grid
was sampled with $n_{\rm r}$ = 1000 radial coordinates, and $n_{\mu}$ = 30
angular coordinates, as discussed in Section~\ref{sec:converge}.

\begin{table}
\centering
\begin{minipage}[c]{0.5\textwidth}
    \caption{\textsc{NFW ionization correction coefficients}}
    \begin{tabular}{lcc}
    \hline
    \hline
    \multicolumn{3}{c}{C\,\textsc{iii}/C\,\textsc{ii}}\\
    \hline
    $n$ & $m$ & \\\cmidrule{2-3}
            & $0$ & $1$ \\
    \hline
    $0$& $48.26$ & $-156.23$ \\
    $1$& $-4.5054$ & $15.377$ \\
    $2$& $0.0850$ & $-0.37572$ \\
    \hline
    \hline
    \multicolumn{3}{c}{N\,\textsc{ii}/N\,\textsc{i}}\\
    \hline
    $n$ & $m$ & \\\cmidrule{2-3}
        & $0$ & $1$ \\
    \hline
    $0$ & $262.58$ & $-55.013$ \\
    $1$ & $-27.121$ & $4.7885$ \\
    $2$ & $0.67994$ & $-0.0980$ \\
    \hline
    \hline
    \multicolumn{3}{c}{Si\,\textsc{iii}/Si\,\textsc{ii}}\\
    \hline
    $n$ & $m$ & \\\cmidrule{2-3}
        & $0$ & $1$ \\
    \hline
    $0$ & $74.500$ & $-155.98$ \\
    $1$ & $-7.1525$ & $15.266$ \\
    $2$ & $0.15165$ & $-0.3709$ \\
    \hline
    \end{tabular}
    \label{tab:NFWcoeff}
\end{minipage}
\end{table}

We began our calculations with a halo mass $\log_{10}M_{200}/M_{\odot}=7.0$,
for each grid value of the radiation intensity and $f_{200}$. As discussed in
Section~\ref{sec:nummeth}, the first iteration of our numerical solution was
initialised with a gas temperature of 20,000~K, and the gas was assumed to be
mostly ionized. No dark matter halo with $\log_{10}M_{200}/M_{\odot}=7.0$ was able to host
gas that would appear as a DLA at $z\sim3$. We then increased the virial mass
of each halo by $0.1~{\rm dex}$, using
as input the final thermal and ionization structure that were output by the
previous model. By initialising the simulations with the solution from the previous
model, the computational requirement was greatly reduced.
We sequentially increased the dark matter halo mass until the peak $N$(\HI) of a
halo had exceeded $\sim10^{21.5}~{\rm cm}^{-2}$. At these \HI\ column densities,
molecular H formation (and presumably star formation) is an important factor to consider
\citep{NotPetSri15}, and is not included in our modelling procedure.

Our final model suite comprises all sightlines through these dark matter halos that
produce an \HI\ column density in excess of $10^{19}~{\rm cm}^{-2}$. In the right
panel of Fig.~\ref{fig:ppic} we show the results of our NFW calculations, which are
qualitatively similar to the plane parallel geometry; the D/H ionization
correction is most sensitive to the column density ratio of the successive stages of
metal ionization (and to a lesser extent on the \HI\ column density). However, the
ionization correction in the NFW geometry also depends on the grid
of parameter values (halo mass, $f_{200}$, etc.), which introduce an uncertainty
in this correction.
The N ion ratio provides the most reliable correction, accurate to within
$2\times10^{-5}$~dex (6~per~cent) for log~$N$(\HI)/cm$^{-2}>20.0$ over the entire grid of parameter
space considered here.
The uncertainty is considerably worse for the C and Si ion ratios, which have
a 30 per cent uncertainty in the correction over the entire range (i.e. a correction
uncertainty of up to $1\times10^{-4}$~dex when the ion ratio is near unity). For each of these
ion ratios we provide fitting formulae for the ionization correction factors,
in the form of Eq.~\ref{eqn:icpoly},
with the $a_{nm}$ coefficients listed in Table~\ref{tab:NFWcoeff}. These
calculations represent the central values of the ionization correction, and
should be used together with the percentage uncertainties quoted above.

\section{Discussion}
\label{sec:disc}

Our calculations with both geometries suggest that deuterium is slightly
underionized relative to hydrogen in metal-poor DLAs. By disabling the
individual physical processes in our code, we conclude that the dominant
physical process that sets the deuterium ionization correction is D~$\leftrightarrow$~H
charge transfer; specifically, the negative ionization correction term is the
result of the preferred (exothermic) charge transfer recombination
reaction D$^{+}$~+~H$^{0}~\to~$D$^{0}$~+~H$^{+}$. Photoionization plays
a minimal role; thus, the most important factor to determine the deuterium
ionization correction in DLAs and sub-DLAs is the H volume density and
the gas temperature, not the incident radiation field.

There are a few remarkable similarities and differences between the two
geometries considered in this paper. The first major difference between the
NFW and plane parallel models is the spread of ion ratios for a given
\HI\ column density; the NFW models exhibit a very small range
in the ion ratios (typically $0.5$~dex), whereas the plane parallel models
cover a much larger range (at least $5$~dex).
This difference is purely a geometrical effect, since the gas density
profile in the NFW geometry is more limited, yet has the advantage
of being more physically motivated. Another difference
between the two geometries is the value of the deuterium ionization
correction using the C and Si ion ratios. The plane parallel models
underpredict the ionization correction for \HI\ column densities
$\lesssim10^{20}$~cm$^{-2}$, and overpredict the correction when
$N$(\HI)~$\gtrsim10^{20}$~cm$^{-2}$. The \NII/\NI\ ion ratio, on the other hand,
is nearly identical regardless of the gas geometry, and therefore
offers the least geometry dependent correction.

The invariance of the \NII/\NI\ ratio with geometry or any of the parameter
space that we have explored is because the charge transfer recombination
reaction N$^{+}$~+~H$^{0}~\to~$N$^{0}$~+~H$^{+}$ sets the \NII/\NI\ ratio
when log\,$N$(\HI)/${\rm cm}^{-2}\gtrsim19.5$. Therefore,
$N$(\NII)/$N$(\NI)~$\propto$~$N$(\DII)/$N$(\DI)~$\propto$~$N$(\HII)/$N$(\HI)
for large \HI\ column densities.
On the other hand, when log\,$N$(\HI)/${\rm cm}^{-2}\lesssim19.5$, primary
photoionization drives the determination of the \NII/\NI\ ratio.
We therefore conclude that the most suitable systems for measuring
the primordial deuterium abundance are those with
log\,$N$(\HI)/${\rm cm}^{-2}\gtrsim20.0$ for the following reasons:
(1) The deuterium ionization correction factor is relatively small (typically $\lesssim0.0005$~dex);
(2) If an ionization correction needs to be applied in the future, the \NII/\NI\ ratio
should depend only on charge exchange, similarly for the \DII/\DI\ ratio,
and should be largely independent of the unknown physical properties of the
system (e.g. density, shape/intensity of the incident radiation field); and
(3) As discussed in \citet{Coo14}, the Lorentzian damping wings
of the \HI~\Lya\ absorption line, together with the multitude of weak high order
Lyman series \DI\ absorption lines, offer an determination of the $N$(\DI)\,/\,$N$(\HI)
ratio that is largely independent of the cloud model.

Under the assumption that D~$\leftrightarrow$~H charge transfer is
the dominant process that sets the D/H ionization correction, we can
combine Eq.~\ref{eqn:dctr} and Eq.~\ref{eqn:ionbalance} to obtain
the following relation:
\begin{equation}
\label{eqn:DHcloudy}
n({\rm D}^{0}) = \frac{n({\rm H}^{0})\,n({\rm H})\,10^{\rm (D/H)_{\rm P}}}{n({\rm H}^{0})+n({\rm H}^{+})\exp(-42.915/T)}
\end{equation}
where $({\rm D/H})_{\rm P}=-4.60$ is the assumed primordial abundance of deuterium.
Eq.~\ref{eqn:DHcloudy} can be applied in post-processing to more
detailed photoionization calculations, such as \textit{Cloudy}, that
include a much larger range of physical processes than our
calculations. Since \NII/\NI\ is independent of geometry, the
relationship between \NII/\NI\ and IC(D/H) using the \textit{Cloudy}
photoionization software and Eq.~\ref{eqn:DHcloudy} should
offer the most reliable ionization correction. We therefore conclude
that the following simple relation should be used to estimate the
deuterium ionization correction factor:
\begin{equation}
\label{eqn:icpoly}
{\rm IC(D/H)} = -\exp(-8.2+1.2\,{\rm IR})
\end{equation}
where IR~$=\log_{10}N({\rm N\,\textsc{ii}})/N({\rm N\,\textsc{i}})$ and,
as discussed in Section~\ref{sec:nfwmod}, IC(D/H) should be used
with a $2\times10^{-5}$~dex (i.e. 6~per~cent) uncertainty in
the ionization correction factor.

Although the deuterium ionization correction estimated herein does not significantly alter
the current determination of the primordial deuterium abundance, it will become important
to consider this systematic offset in the near-future. We propose that the \NII/\NI\
ratio provides the most robust ionization correction for deuterium, since it is
insensitive to the various known and unknown physical properties of
metal-poor DLAs (e.g. cloud geometry, halo mass, \HI\ column density, etc.).
Unfortunately, nitrogen is notably underabundant in metal-poor DLAs
\citep{Pet08,Coo11,PetCoo12b,Zaf14}; there are typically just three nitrogen atoms
for every 10,000 D atoms in the most metal-poor systems. 
Nevertheless, measuring the N ion ratio should be a goal for future
high precision surveys that aim to accurately measure
the primordial abundance of deuterium in the most metal-poor DLAs.


\section{Summary and Conclusions}
\label{sec:conc}

The $N$(\DI)\,/\,$N$(\HI) column density ratio that is measured using
quasar absorption line systems
is generally assumed to be equal to the deuterium abundance (D/H).
We have tested this assumption by developing a software package that 
calculates the relative ionization of deuterium, hydrogen, and a selection
of the most abundant metal ions. Our computations are similar
in spirit to the commonly used photoionization software \textit{Cloudy}.
The main advantage of our code, aside from including the atomic data
for deuterium, is that we are able to model the density, thermal, and
ionization profile of gas held in hydrostatic equilibrium with an
NFW dark matter potential. On the basis of our calculations,
we draw the following conclusions:\\

\noindent ~~(i) For the simple case of a uniform density, plane parallel
slab of gas with similar properties to DLAs, we have confirmed that our
code produces comparable results as \textit{Cloudy} for the ions
that are used in our work.

\smallskip

\noindent ~~(ii) We performed a suite of calculations using
a plane parallel geometry to estimate the correction that must be
applied to the measured column density ratio $N$(\DI)/$N$(\HI) to determine
the D/H abundance. This correction is most sensitive to the relative
column densities of successive stages of metal ionization and, to a
lesser degree, the total \HI\ column density.

\smallskip

\noindent ~~(iii) We also calculated the deuterium ionization
correction for a more realistic gas density distribution, whereby
a gas cloud is confined to the potential of an NFW dark
matter halo irradiated by the \citet{HarMad12} background.
For both the plane parallel and NFW geometry,
we provide fitting functions to estimate the deuterium ionization
correction using either the \CIII/\CII, \NII/\NI, or \SiIII/\SiII\ ion ratios
and the \HI\ column density.

\smallskip

\noindent ~~(iv) The relationship between the \NII/\NI\ ion ratio
and the deuterium ionization correction is remarkable. This
relationship is essentially independent of all other physical
properties of our model, and provides the most accurate and
reliable deuterium ionization correction. The \CIII/\CII\ and
\SiIII/\SiII\ ion ratios, although more observationally accessible,
are more dependent on both the physical conditions and the
geometry of the gas.

\smallskip

\noindent ~~(v) We propose that systems with log\,$N$(\HI)/${\rm cm}^{-2}\gtrsim20.0$
are the most suitable systems to measure the primordial D/H ratio. In
this \HI\ column density regime, charge exchange ensures that
$N$(\NII)/$N$(\NI)~$\propto$~$N$(\DII)/$N$(\DI)~$\propto$~$N$(\HII)/$N$(\HI),
allowing a reliable ionization correction factor to be determined.

\smallskip

\noindent ~~(vi) Our work provides the first quantitative analysis
of the deuterium ionization correction in metal-poor DLAs, and
confirms the qualitative conclusions drawn by \citet{Sav02}:
The deuterium ionization correction can be safely neglected
when the \DI/\HI\ abundance ratio is measured in a DLA that
is in thermal and ionization equilibrium.

The deuterium ionization correction for a typical system in the
recent analysis by \citet{Coo14} is likely to be $\lesssim0.1$
per cent, which is well below the current measurement
uncertainty (2 per cent); the ionization correction is therefore
not an important consideration at present. However, new cases
of metal-poor DLAs that exhibit clean \DI\ absorption lines
are being identified, and the search for such systems will be
greatly expanded with the next generation of 30+\,m telescopes
coming on line in the next decade. At that time, a reliable
ionization correction factor for deuterium will be required,
as the statistics of D/H measurements improve significantly
beyond the current limited dataset.

\section*{Acknowledgements}
We thank P.~Madau and J.~X.~Prochaska for useful
discussions about the work described in this paper.
We also thank an anonymous referee for offering
many constructive comments that helped to improve
the presentation of our work.
R.~J.~C. is currently supported by NASA through
Hubble Fellowship grant HST-HF-51338.001-A, awarded by the
Space Telescope Science Institute, which is operated by the
Association of Universities for Research in Astronomy, Inc.,
for NASA, under contract NAS5- 26555.





\label{lastpage}

\end{document}